\def\tr#1{{\rm tr} #1}
\def\R{R}

\documentstyle{livrev}
\begin{document}

\title{Local and global existence theorems for the Einstein equations}
\author{Alan D. Rendall, Max-Planck-Institut f\"ur Gravitationsphysik,
\\Am M\"uhlenberg 1, 14476 Golm, Germany, 
\\rendall@aei-potsdam.mpg.de}
\date{}
\maketitle

\begin{abstract}
This article is a guide to the literature on existence theorems for the
Einstein equations which also draws attention to open problems in the
field. The local in time Cauchy problem, which is relatively well
understood, is treated first. Next global results for solutions with 
symmetry are discussed. A selection of results from Newtonian theory
and special relativity which offer useful comparisons is presented.
This is followed by a survey of global results in the case of small data
and results on constructing spacetimes with given singularity structure.
The article ends with some miscellaneous topics connected with the main 
theme.
\end{abstract}

\section{Introduction}\label{introduction}

Many of the mathematical models occurring in physics involve systems of
partial differential equations. Only rarely can these equations be 
solved by explicit formulae. When they cannot, physicists frequently
resort to approximations. There is, however, another approach which
is complementary. This consists in determining the qualitative
behaviour of solutions, without knowing them explicitly. The first
and most fundamental step in doing this is to establish the existence
of solutions under appropriate circumstances. Unfortunately, this is
often hard, and obstructs the way to obtaining more interesting
information. It may appear to the outside observer that existence
theorems become a goal in themselves to some researchers. It is 
important to remember that, from a more general point of view, they
are only a first step.

The basic partial differential equations of general relativity are
Einstein's equations. In general they are coupled to other partial
differential equations describing the matter content of spacetime.
The Einstein equations are essentially hyperbolic in nature. In
other words, the general properties of solutions are similar to those
found for the wave equation. It follows that it is reasonable to try to
determine a solution by initial data on a spacelike hypersurface.
Thus the Cauchy problem is the natural context for existence theorems
for the Einstein equations. The Einstein equations are also nonlinear.
This means that there is a big difference between the local and global 
Cauchy problems. A solution evolving from regular data may develop 
singularities.

A special feature of the Einstein equations is that they are
diffeomorphism invariant. If the equations are written down in
an arbitrary coordinate system then the solutions of these coordinate
equations are not uniquely determined by initial data. Applying a
diffeomorphism to one solution gives another solution. If this 
diffeomorphism is the identity on the chosen Cauchy surface up to
first order then the data are left unchanged by this transformation.
In order to obtain a system for which uniqueness in the Cauchy problem
holds in the straightforward sense it does for the wave equation, some 
coordinate or gauge fixing must be carried out.

Another special feature of the Einstein equations is that initial 
data cannot be given freely. They must satisfy constraint
equations. To prove the existence of a solution of the Einstein
equations, it is first necessary to prove the existence of a
solution of the constraints. The usual method of solving the
constraints relies on the theory of elliptic equations.

The local existence theory of solutions of the Einstein equations
is rather well understood. Section \ref{local} points out some of the things 
which are not known. On the other hand, the problem of proving general 
global existence theorems for the Einstein equations is beyond the reach of 
the mathematics presently available. To make some progress, it is necessary 
to concentrate on simplified models. The most common simplifications are
to look at solutions with various types of symmetry and solutions for
small data. These two approaches are reviewed in sections \ref{symmetric}
and \ref{small} respectively. A different approach is to prove the 
existence of solutions with a prescribed singularity structure. This is
discussed in section \ref{prescribe}. Section \ref{further} collects some 
miscellaneous results which cannot easily be classified. With the motivation
that insights about the properties of solutions of the Einstein equations
can be obtained from the comparison with Newtonian theory and special
relativity, relevant results from those areas are presented in section
\ref{newtonian}.     

The area of research reviewed in the following relies heavily on the theory
of differential equations, particularly that of hyperbolic partial 
differential equations. For the benefit of readers with little background
in differential equations, some general references which the author has
found to be useful will be listed. A thorough introduction to ordinary
differential equations is given in \cite{hartman82}. A lot of intuition
for ordinary differential equations can be obtained from\cite{hubbard91}.
The article\cite{arnold88} is full of information, in rather compressed
form. A classic introductory text on partial differential equations, where
hyperbolic equations are well represented, is \cite{john82}. Useful texts
on hyperbolic equations, some of which explicitly deal with the Einstein
equations, are \cite{taylor96,kichenassamy96a,racke92,majda84,strauss89,john90,
evans98} 

An important aspect of existence theorems in general relativity which
one should be aware of is their relation to the cosmic censorship 
hypothesis. This point of view was introduced in an influential paper by
Moncrief and Eardley\cite{moncrief81a}. An extended discussion of the
idea can be found in\cite{chrusciel91a}  

\section{Local existence}\label{local}
\subsection{The constraints}\label{constraints}

The unknowns in the constraint equations are the initial data for the 
Einstein equations. These consist of a three-dimensional manifold $S$,
a Riemannian metric $h_{ab}$ and a symmetric tensor $k_{ab}$ on $S$,
and initial data for any matter fields present. The equations are: 
\begin{eqnarray}
&R-k_{ab}k^{ab}+(h^{ab}k_{ab})^2=16\pi\rho         \\ 
&\nabla^a k_{ab}-\nabla_b(h^{ac}k_{ac})=8\pi j_b   
\end{eqnarray}
Here $R$ is the scalar curvature of the metric $h_{ab}$ and $\rho$ and $j_a$
are projections of the energy-momentum tensor. Assuming matter fields which
satisfy the dominant energy condition implies that $\rho\ge (j_aj^a)^{1/2}$.
This means that the trivial procedure of making an arbitrary choice of 
$h_{ab}$ and $k_{ab}$ and defining $\rho$ and $j_a$ by equations (1) and (2) 
is of no use for producing physically interesting solutions. 

The usual method for solving the equations (1) and (2) is the conformal 
method\cite{choquet80}. In this
method parts of the data (the so-called free data) are chosen, and the
constraints imply four elliptic equations for the remaining parts. The
case which has been studied most is the constant mean curvature (CMC)
case, where $\tr k=h^{ab}k_{ab}$ is constant. In that case there is an
important simplification. Three of the elliptic equations, which form
a linear system, decouple from the remaining one. This last equation,
which is nonlinear, but scalar, is called the Lichnerowicz equation.
The heart of the existence theory for the constraints in the CMC case
is the theory of the Lichnerowicz equation.

Solving an elliptic equation is a non-local problem and so boundary
conditions or asymptotic conditions are important. For the constraints
the cases most frequently considered in the literature are that where
$S$ is compact (so that no boundary conditions are needed) and that
where the free data satisfy some asymptotic flatness conditions. In
the CMC case the problem is well understood for both kinds of boundary 
conditions\cite{cantor79,christodoulou81,isenberg95}. The other case which 
has been studied in detail is that of hyperboloidal data \cite{andersson92}. 
The kind of theorem which is obtained is that sufficiently differentiable 
free data, in some cases required to satisfy some global restrictions, can be 
completed in a unique way to a solution of the constraints. 

In the non-CMC case our understanding is much more limited although
some results have been obtained in recent years (see \cite{isenberg96,
choquet99} and references therein.) It is an 
important open problem to extend these so that an overview is obtained
comparable to that available in the CMC case. Progress on this could 
also lead to a better understanding of the question, when a spacetime
which admits a compact, or asymptotically flat, Cauchy surface also
admits one of constant mean curvature. Up to now there are only
isolated examples which exhibit obstructions to the existence of CMC
hypersurfaces\cite{bartnik88a}.

It would be interesting to know whether there is a useful concept of
the most general physically reasonable solutions of the constraints
representing regular initial configurations. Data of this kind should not 
themselves contain singularities. Thus it seems reasonable to suppose at
least that the metric $h_{ab}$ is complete and that the length of $k_{ab}$, 
as measured using $h_{ab}$, is bounded. Does the existence of solutions of 
the constraints imply a restriction on the topology of $S$ or on the 
asymptotic geometry of the data? This question is largely open, and it seems
that information is available only in the compact and asymptotically flat 
cases. In the case of compact $S$, where there is no asymptotic regime,
there is known to be no topological restriction. In the asymptotically flat 
case there is also no topological restriction implied by the constraints
beyond that implied by the condition of asymptotic flatness itself\cite{witt86}
This shows in particular that any manifold which is obtained by deleting a
point from a compact manifold admits a solution of the constraints satisfying
the minimal conditions demanded above. A starting point for going beyond
this could be the study of data which are asymptotically homogeneous. For 
instance, the Schwarzschild solution contains interesting CMC hypersurfaces 
which are asymptotic to the product of a 2-sphere with the real line. More 
general data of this kind could be useful for the study of the dynamics of 
black hole interiors\cite{rendall96a}

To sum up, the conformal approach to solving the constraints, which is
the standard one up to now, is well understood in the compact, asymptotically 
flat and hyperboloidal cases under the constant mean curvature assumption,
and only in these cases. For some other approaches see \cite{bartnik93a},
\cite{bartnik93b} and \cite{york99}.

\subsection{The vacuum evolution equations}\label{vacuum}

The main aspects of the local in time existence theory for the Einstein
equations can be illustrated by restricting to smooth (i.e. infinitely 
differentiable) data for the vacuum Einstein equations. The generalizations
to less smooth data and matter fields are discussed in sections  
\ref{differentiability} and \ref{matter} respectively. In the vacuum 
case the data are $h_{ab}$ and $k_{ab}$ on a three-dimensional manifold $S$, 
as discussed in section \ref{constraints}.  
A solution corresponding to these data is given by a 
four-dimensional manifold $M$, a Lorentz metric $g_{\alpha\beta}$ on
$M$ and an embedding of $S$ in $M$. Here $g_{\alpha\beta}$ is supposed to
be a solution of the vacuum Einstein equations while $h_{ab}$ and $k_{ab}$
are the induced metric and second fundamental form of the embedding,
respectively.

The basic local existence theorem says that, given smooth data for the 
vacuum Einstein equations, there exists a smooth solution of the equations
which gives rise to these data\cite{choquet80}. Moreover, it can be assumed 
that the image of $S$ under the given embedding is a Cauchy surface for the 
metric $g_{\alpha\beta}$. The latter fact may be expressed loosely, 
identifying $S$ with its image, by the statement that $S$ is a Cauchy 
surface. A solution of the Einstein equations with given initial data having 
$S$ as a Cauchy surface is called a Cauchy development of those data. The 
existence theorem is local because it says nothing about the size of the 
solution obtained. A Cauchy development of given data has many open subsets 
which are also Cauchy developments of that data.

It is intuitively clear what it means for one Cauchy development to be
an extension of another. The extension is called proper if it is strictly
larger than the other development. A Cauchy development which has no proper
extension is called maximal. The standard global uniqueness theorem for
the Einstein equations uses the notion of the maximal development. It is
due to Choquet-Bruhat and Geroch\cite{choquet69}. It says that the maximal 
development of any Cauchy data is unique up to a diffeomorphism which fixes 
the initial hypersurface. It is also possible to make a statement of Cauchy 
stability which says that, in an appropriate sense, the solution depends 
continuously on the initial data. Details on this can be found in
\cite{choquet80}.

A somewhat stronger form of the local existence theorem is to say that
the solution exists on a uniform time interval in all of space. The 
meaning of this is not a priori clear, due to the lack of a preferred 
time coordinate in general relativity. The following is a formulation which 
is independent of coordinates. Let $p$ be a point of $S$. The temporal
extent $T(p)$ of a development of data on $S$ is the supremum of the length 
of all causal curves in the development passing through $p$. In this way a
development defines a function $T$ on $S$. The development can be regarded
as a solution which exists on a uniform time interval if $T$ is bounded
below by a strictly positive constant. For compact $S$ this is a 
straightforward consequence of Cauchy stability. In the case of asymptotically 
flat data it is less trivial. In the case of the vacuum Einstein equations it 
is true, and in fact the function $T$ grows at least linearly at infinity
\cite{christodoulou81}.  

When proving the above local existence and global uniqueness theorems it
is necessary to use some coordinate or gauge conditions. At least no
explicitly diffeomorphism-invariant proofs have been found up to now.
Introducing these extra elements leads to a system of reduced equations,
whose solutions are determined uniquely by initial data in the strict 
sense, and not just uniquely up to diffeomorphisms. When a solution of the
reduced equations has been obtained, it must be checked that it is a 
solution of the original equations. This means checking that the constraints
and gauge conditions propagate. There are many methods for reducing the
equations. An overview of the possibilities may be found in \cite{friedrich96}

\subsection{Questions of differentiability}\label{differentiability}

Solving the Cauchy problem for a system of partial differential equations
involves specifying a set of initial data to be considered, and determining
the differentiability properties of solutions. Thus two regularity properties
are involved - the differentiability of the allowed data, and that of the
corresponding solutions. Normally it is stated that for all data 
with a given regularity, solutions with a certain type of regularity
are obtained. For instance in the section \ref{vacuum} we chose both types of
regularity to be \lq infinitely differentiable\rq. The correspondence
between the regularity of data and that of solutions is not a matter of
free choice. It is determined by the equations themselves, and in
general the possibilities are severely limited. A similar issue arises
in the context of the Einstein constraints, where there is a correspondence
between the regularity of free data and full data.

The kinds of regularity properties which can be dealt with in the Cauchy
problem depends of course on the mathematical techniques available.
When solving the Cauchy problem for the Einstein equations it is necessary
to deal at least with nonlinear systems of hyperbolic equations. (There
may be other types of equations involved, but they will be ignored here.)
For general nonlinear systems of hyperbolic equations there is essentially
only one technique known, the method of energy estimates. This method is
closely connected with Sobolev spaces, which will now be discussed briefly.

Let $u$ be a real-valued function on ${\bf R}^n$. Let:
$$\|u\|_s=(\sum_{i=0}^s\int |D^i u|^2 (x) dx)^{1/2}$$
The space of functions for which this quantity is finite is the Sobolev
space $H^s({\bf R}^n)$. Here $|D^i u|^2$ denotes the sum of the squares of 
all partial derivatives of $u$ of order $i$. Thus the Sobolev space
$H^s$ is the space of functions, all of whose partial derivatives up
to order $s$ are square integrable. Similar spaces can be defined for
vector valued functions by taking a sum of contributions from the
separate components in the integral. It is also possible to define 
Sobolev spaces on any Riemannian manifold, using covariant derivatives.
General information on this can be found in \cite{aubin82}. Consider now a
solution $u$ of the wave equation in Minkowski space. Let $u(t)$ be
the restriction of this function to a time slice. Then it is easy to
compute that, provided $u$ is smooth and $u(t)$ has compact support for 
each $t$, the quantity $\|Du(t)\|^2_s+\|\partial_t u(t)\|^2_s$ is time 
independent for each $s$. For $s=0$ this is just the energy of a solution of 
the wave equation. For a general nonlinear hyperbolic system, the Sobolev 
norms are no longer time-independent. The constancy in time is replaced by 
certain inequalities. Due to the similarity to the energy for the wave 
equation, these are called energy estimates. They constitute the foundation 
of the theory of hyperbolic equations. It is because of these estimates that
Sobolev spaces are natural spaces of initial data in the Cauchy problem
for hyperbolic equations. The energy estimates ensure that a solution
evolving from data belonging to a given Sobolev space on one spacelike
hypersurface will induce data belonging to the same Sobolev space on later
spacelike hypersurfaces. In other words, the property of belonging to a
Sobolev space is propagated by the equations. Due to the locality properties 
of hyperbolic equations (existence of a finite domain of dependence), it is 
useful to introduce the spaces $H^s_{\rm loc}$ which are defined by the 
condition that whenever the domain of integration is restricted to a compact 
set the integral defining the space $H^s$ is finite.

In the end the solution of the Cauchy problem should be a function
which is differentiable enough in order that all derivatives 
which occur in the equation exist in the usual (pointwise) sense.
A square integrable function is in general defined only almost everywhere
and the derivatives in the above formula must be interpreted as
distributional derivatives. For this reason a connection between Sobolev 
spaces and functions whose derivatives exist pointwise is required. This is 
provided by the Sobolev embedding theorem. This says that if a function $u$ 
on ${\bf R}^n$ belongs to the Sobolev space $H^s_{\rm loc}$ and if $k<s-n/2$ 
then there is a $k$ times continuously differentiable function which agrees 
with $u$ except on a set of measure zero.

In the existence and uniqueness theorems stated in section \ref{vacuum}, the
assumptions on the initial data for the vacuum Einstein equations can be 
weakened to say that $h_{ab}$ should belong to $H^s_{\rm loc}$ and $k_{ab}$ 
to $H^{s-1}_{\rm loc}$. Then, provided $s$ is large enough, a solution is 
obtained which belongs to $H^s_{\rm loc}$. In fact its restriction to
any spacelike hypersurface also belongs to $H^s_{\rm loc}$, a property
which is a priori stronger. The details of how large $s$ must be would be out 
of place here, since they involve examining the detailed structure of the 
energy estimates. However there is a simple rule for computing the required 
value of $s$. The value of $s$ needed to obtain an existence theorem for the 
Einstein equations is that for which the Sobolev embedding theorem, applied to
spatial slices, just ensures that the metric is continuously differentiable.
Thus the requirement is that $s>n/2+1=5/2$, since $n=3$. It follows that the 
smallest possible integer $s$ is three. Strangely enough, uniqueness up to 
diffeomorphisms is only known to hold for $s\ge 4$. The reason is that in 
proving the uniqueness theorem a diffeomorphism must be carried out, which 
need not be smooth. This apparently leads to a loss of one derivative.
It would be desirable to show that uniqueness holds for $s=3$ and to
close this gap, which has existed for many years. There exists a definition
of Sobolev spaces for an arbitrary real number $s$, and hyperbolic 
equations can also be solved in the spaces with $s$ not an integer 
\cite{taylor91}. Presumably these techniques could be applied to prove local 
existence for the Einstein equations with $s$ any real number greater than 
$5/2$. However this has apparently not been done explicitly in the literature.

Consider now $C^\infty$ initial data. Corresponding to these data there is
a development of class $H^s$ for each $s$. It could conceivably be the case
that the size of these developments shrinks with increasing $s$. In that 
case their intersection might contain no open neighbourhood of the initial
hypersuface, and no smooth development would be obtained. Fortunately it
is known that the $H^s$ developments cannot shrink with increasing $s$,
and so the existence of a $C^\infty$ solution is obtained for $C^\infty$
data. It appears that the $H^s$ spaces with $s>5/2$ are the only spaces
containing the space of smooth functions for which it has been proved that
the Einstein  equations are locally solvable.

What is the motivation for considering regularity conditions other than
the apparently very natural $C^\infty$ condition?  One motivation concerns
matter fields and will be discussed in section \ref{matter}. Another is 
the idea that assuming the existence of many derivatives which have no direct
physical significance seems like an admission that the problem has not
been fully understood. A further reason for considering low regularity
solutions is connected to the possibility of extending a local existence
result to a global one. If the proof of a local existence theorem is
examined closely it is generally possible to give a continuation criterion.
This is a statement that if a solution on a finite time interval is such 
that a certain quantity constructed from the solution is bounded on that 
interval, then the solution can be extended to a longer time interval. (In
applying this to the Einstein equations we need to worry about introducing 
an appropriate time coordinate.) If it can be shown that the relevant 
quantity is bounded on any finite time interval where a solution exists, 
then global existence follows. It suffices to consider the maximal interval 
on which a solution is defined, and obtain a contradiction if that interval
is finite. This description is a little vague, but contains the essence of
a type of argument which is often used in global existence proofs. The
problem in putting it into practise is that often the quantity whose
boundedness has to be checked contains many derivatives, and is therefore
difficult to control. If the continuation criterion can be improved by
reducing the number of derivatives required, then this can be a significant
step towards a global result. Reducing the number of derivatives in the
continuation criterion is closely related to reducing the number of 
derivatives of the data required for a local existence proof.

A striking example is provided by the work of Klainerman and Machedon
\cite{klainerman95} on the Yang-Mills equations in Minkowski space. Global 
existence in this case was first proved by Eardley and Moncrief
\cite{eardley82}, assuming initial data of
sufficiently high differentiability. Klainerman and Machedon gave a new 
proof of this which, though technically complicated, is based on a
conceptually simple idea. They prove a local existence theorem for data 
of finite energy. Since energy is conserved this immediately proves 
global existence. In this case finite energy corresponds to the Sobolev
space $H^1$ for the gauge potential. Of course a result of this kind 
cannot be expected for the Einstein equations, since spacetime singularities
do sometimes develop from regular initial data. However, some weaker 
analogue of the result could exist. 

\subsection{Matter fields}\label{matter}
Analogues of the results for the vacuum Einstein equations given in section
\ref{vacuum}
are known for the Einstein equations coupled to many types of matter
model. These include perfect fluids, elasticity theory, kinetic theory,
scalar fields, Maxwell fields, Yang-Mills fields and combinations of these.
An important restriction is that the general results for perfect fluids
and elasticity apply only to situations where the energy density is 
uniformly bounded away from zero on the region of interest. In particular
they do not apply to cases representing material bodies surrounded by 
vacuum. In cases where the energy density, while everywhere positive,
tends to zero at infinity, a local solution is known to exist, but it is
not clear whether a local existence theorem can be obtained which is 
uniform in time. In cases where the fluid has a sharp boundary,
ignoring the boundary leads to solutions of the Einstein-Euler equations 
with low differentiability (cf. section \ref{differentiability}), while taking
it into account explicitly leads to a free boundary problem. This will be 
discussed in more detail in section \ref{freeboundary}.
In the case of kinetic or field theoretic matter models it makes no 
difference whether the energy density vanishes somewhere or not. 

\subsection{Free boundary problems}\label{freeboundary}

In applying general relativity one would like to have solutions of the
Einstein-matter equations modelling material bodies. As will be discussed
in section \ref{stationary} there are solutions available for describing
equilibrium situations. However dynamical situations require solving a free
boundary problem if the body is to be made of fluid or an elastic solid.
We will now discuss the few results which are known on this subject. For
a spherically symmetric self-gravitating fluid body in general relativity
a local in time existence theorem was proved in \cite{kind93}. This 
concerned the case where the density of the fluid at the boundary is
non-zero. In \cite{rendall92b} a local existence theorem was proved for certain
equations of state with vanishing boundary density. These solutions need
not have any symmetry but they are very special in other ways. In particular
they do not include small perturbations of the stationary solutions
discussed in section \ref{stationary}. There is no general result on this
problem up to now.

Remarkably, the free boundary problem for a fluid body is also poorly
understood in classical physics. There is a result for a viscous fluid
\cite{secchi91} but in the case of a perfect fluid the problem was wide 
open until very recently. Now a major step forward has been taken by 
Wu\cite{wu99}, who obtained a result for a fluid which is incompressible
and irrotational. There is a good physical reason why local existence for
a fluid with a free boundary might fail. This is the Rayleigh-Taylor
instability which involves perturbations of fluid interfaces which grow
with unbounded exponential rates. (Cf. the discussion in \cite{beale93}.)
It turns out that in the case considered by Wu this instability does not 
cause problems and there is no reason to expect that a self-gravitating
compressible fluid with rotation in general relativity with a free 
boundary cannot also be described by a well-posed free boundary value 
problem.

One of the problems in tackling the initial value problem for
a dynamical fluid body is that the boundary is moving. It would be very
convenient to use Lagrangian coordinates, since in those coordinates the
boundary is fixed. Unfortunately, it is not at all obvious that the Euler
equations in Lagrangian coordinates have a well-posed initial value problem,
even in the absence of a boundary. It was, however, recently shown by 
Friedrich\cite{friedrich98b} that it is possible to treat the Cauchy problem 
for fluids in general relativity in Lagrangian coordinates.

\section{Global symmetric solutions}\label{symmetric}
\subsection{Stationary solutions}\label{stationary}

Many of the results on global solutions of the Einstein equations involve
considering classes of spacetimes with Killing vectors. A particularly
simple case is that of a timelike Killing vector, i.e. the case of 
stationary spacetimes. In the vacuum case there are very few solutions
satisfying physically reasonable boundary conditions. This is related
to no hair theorems for black holes and lies outside the scope of this 
review. More information on the topic can be found in the book of Heusler
\cite{heusler96} and in his Living Review\cite{heusler98}.
The case of phenomenological matter models has been reviewed
in \cite{rendall97c}. The account given there will be updated in
the following. 

The area of stationary solutions of the Einstein equations coupled to
field theoretic matter models has been active in recent years as a 
consequence of the discovery by Bartnik and McKinnon\cite{bartnik88b} of a 
discrete family of regular static spherically symmetric solutions of the 
Einstein-Yang-Mills equations with gauge group $SU(2)$. The equations
to be solved are ordinary differential equations and in \cite{bartnik88b}
they were solved numerically by a shooting method. The first existence proof 
for a solution of this kind is due to Smoller, Wasserman, Yau and McLeod
\cite{smoller91} and involves an arduous qualitative analysis of the 
differential equations. The work on the Bartnik-McKinnon solutions, including 
the existence theorems, has been extended in many directions. Recently
static solutions of the Einstein-Yang-Mills equations which are not
spherically symmetric were discovered numerically \cite{kleihaus98}. It is a
challenge to prove the existence of solutions of this kind. Now the
ordinary differential equations of the previously known case are replaced
by elliptic equations. Moreover, the solutions appear to still be discrete, 
so that a simple perturbation argument starting from the spherical case does
not seem feasible. In another development it was shown that a linearized
analysis indicates the existence of stationary non-static solutions
\cite{brodbeck97}. It would be desirable to study the question of 
linearization stability in this case, which, if the answer were favourable, 
would give an existence proof for solutions of this kind.

Now we return to phenomenological matter models, starting with the case
of spherically symmetric static solutions. Basic existence theorems for
this case have been proved for perfect fluids\cite{rendall91},
collisionless matter \cite{rein93}, \cite{rein94a} and elastic bodies
\cite{park98}. The last of these is the solution to an open problem 
posed in \cite{rendall97c}. All these theorems demonstrate the existence 
of solutions which are everywhere smooth and exist globally as functions
of area radius for a general class of constitutive relations. The physically 
significant question of the finiteness of the mass of these configurations 
was only answered in these papers under restricted circumstances. For 
instance, in the case of perfect fluids and collisionless matter, solutions 
were constructed by perturbing about the Newtonian case. Solutions for an 
elastic body were obtained by perturbing about the case of isotropic 
pressure, which is equivalent to a fluid. Further progress on the question
of the finiteness of the mass of the solutions was made in the case of a 
fluid by Makino\cite{makino98}, who gave a rather general criterion on the
equation of state ensuring the finiteness of the radius. Makino's criterion 
was generalized to kinetic theory in \cite{rein98a}. This resulted in 
existence proofs for various models which have been considered in galactic 
dynamics and which had previously been constructed numerically. (Cf. 
\cite{binney87}, \cite{shapiro85} for an account of these models in the 
non-relativistic and relativistic cases respectively.) 

In the case of self-gravitating Newtonian spherically symmetric 
configurations of collisionless matter, it can be proved that the phase 
space density of particles depends only on the energy of the particle
and the modulus of its angular momentum\cite{batt86}. This is known as
Jeans' theorem. It was already shown in \cite{rein94a} that the naive
generalization of this to the general relativistic case does not hold
if a black hole is present. Recently counterexamples to the generalization 
of Jeans' theorem to the relativistic case which are not dependent on a
black hole were constructed by Schaeffer\cite{schaeffer99}. It remains to 
be seen whether there might be a natural modification of the formulation 
which would lead to a true statement.

For a perfect fluid there are results stating that a static solution is 
necessarily spherically symmetric\cite{lindblom94}. They still require 
a restriction on the equation of state which it would be desirable to
remove. A similar result is not to be expected in the case of other matter
models, although as yet no examples of non-spherical static solutions are
available. In the Newtonian case examples have been constructed by Rein
\cite{reinu1}. (In that case static solutions are defined to be those where 
the particle current vanishes.) For a fluid there is an existence theorem
for solutions which are stationary but not static (models for rotating
stars)\cite{heilig95}. At present there are no corresponding theorems
for collisionless matter or elastic bodies. In \cite{reinu1} stationary,
non-static configurations of collisionless matter were constructed in the
Newtonian case.

For some remarks on the question of stability see section \ref{hydro}.

\subsection{Spatially homogeneous solutions}\label{homogeneous}

A solution of the Einstein equations is called spatially homogeneous if
there exists a group of symmetries with three-dimensional spacelike orbits.
In this case there are at least three linearly independent spacelike
Killing vector fields. For most matter models the field equations reduce
to ordinary differential equations. (Kinetic matter leads to an
integro-differential equation.) The most important results in this area
have been reviewed in a recent book edited by Wainwright and Ellis\cite{
wainwright97}. See, in particular, part two of the book. There remain a host 
of interesting and accessible open questions. The spatially homogeneous 
solutions have the advantage that it is not necessary to stop at just 
existence theorems; information on the global qualitative behaviour of 
solutions can also be obtained. 

An important question which has been open for a long time concerns the 
mixmaster model, as discussed in \cite{rendall97d}. This is a class
of spatially homogeneous solutions of the vacuum Einstein equations
which are invariant under the group $SU(2)$. A special subclass of these
$SU(2)$-invariant solutions, the (parameter-dependent) Taub-NUT solution, 
is known explicitly in terms of elementary functions. The Taub-NUT solution 
has a simple initial singularity which is in fact a Cauchy horizon. All 
other vacuum solutions admitting a transitive action of $SU(2)$ on 
spacelike hypersurfaces (Bianchi type IX solutions) will be called
generic in the present discussion. These generic Bianchi IX solutions
(which might be said to constitute the mixmaster solution proper) have 
been believed for a long time to have singularities which are oscillatory
in nature where some curvature invariant blows up. This belief was based
on a combination of heuristic considerations and numerical calculations.
Although these together do make a persuasive case for the accepted picture,
until very recently there were no mathematical proofs of the these 
features of the mixmaster model available. This has now changed. 
First, a proof of curvature blow-up and oscillatory behaviour for a
simpler model (a solution of the Einstein-Maxwell equations) which shares 
many qualitative features with the mixmaster model was obtained by 
Weaver\cite{weaver99a}. In the much more difficult case of the mixmaster 
model itself corresponding results were obtained by 
Ringstr\"om\cite{ringstrom99a}. Forthcoming work of Ringstr\"om extends 
this analysis to prove the correctness of other properties of the mixmaster 
model suggested by heuristic and numerical work. 

Ringstr\"om's analysis of the mixmaster model is potentially of great 
significance for the mathematical understanding of singularities of the
Einstein equations in general. Thus its significance goes far beyond the
spatially homogeneous case. According to extensive investigations of
Belinskii, Khalatnikov and Lifshitz (see \cite{lifshitz63}, 
\cite{belinskii70}, \cite{belinskii82} and
references therein) the mixmaster model should provide an approximate 
description for the general behaviour of solutions of the Einstein equations
near singularities. This should apply to many matter models as well as to
the vacuum equations. The work of Belinskii, Khalatnikov and Lifshitz
(BKL) is hard to understand and it is particularly difficult to find a
precise mathematical formulation of their conclusions. This has caused
many people to remain sceptical about the validity of the BKL picture.
Nevertheless, it seems that nothing has ever been found which 
indicates any significant flaws in the final version. As long as the
mixmaster model itself was not understood this represented a fundamental
obstacle to progress on understanding the BKL picture mathematically.
The removal of this barrier opens up an avenue to progress on this issue.

Some recent and qualitatively new results concerning the asymptotic 
behaviour of spatially homogeneous solutions of the Einstein-matter 
equations, both close to the initial singularity and in a phase of 
unlimited expansion, (and with various matter models) can be found in 
\cite{rendall99a} and \cite{wainwright99a}. These show in particular that 
the dynamics can depend sensitively on the form of matter chosen. (Note
that these results are consistent with the BKL picture.)

\subsection{Spherically symmetric solutions}\label{spherical}

The most extensive results on global inhomogeneous solutions of the
Einstein equations obtained up to now concern spherically symmetric
solutions of the Einstein equations coupled to a massless scalar 
field with asymptotically flat initial data.
In a series of papers Christodoulou \cite{christodoulou86a,
christodoulou86b,christodoulou87a,christodoulou87b,christodoulou91,
christodoulou93a,christodoulou94,christodoulou99} has proved a
variety of deep results on the global structure of these solutions.
Particularly notable are his proofs that naked singularities can 
develop from regular initial data \cite{christodoulou94} and that this 
phenomenon is unstable with respect to perturbations of the data 
\cite{christodoulou99}. In related work Christodoulou 
\cite{christodoulou95,christodoulou96a,christodoulou96b} has studied 
global spherically symmetric solutions of the Einstein equations coupled to 
a fluid with a special equation of state (the so-called two-phase model).

The rigorous investigation of the spherically symmetric collapse of 
collisionless matter in general relativity was initiated by Rein and the author
\cite{rein92}, who showed that the evolution of small initial data leads to
geodesically complete spacetimes where the density and curvature fall off 
at large times. Later it was shown\cite{rein95a} that independent of the 
size of the initial data the first singularity, if there is one at all,
must occur at the centre of symmetry. This result uses a time coordinate of
Schwarzschild type; an analogous result for a maximal time coordinate was
proved in \cite{rendall97e}. The question of what happens for general large
initial data could not yet be answered by analytical techniques. In 
\cite{rein98b} numerical methods were applied in order to try to make some 
progress in this direction. The results are discussed in the next paragraph.

Despite the range and diversity of the results obtained by Christodoulou
on the spherical collapse of a scalar field, they do not encompass some
of the most interesting phenomena which have been observed numerically.
These are related to the issue of critical collapse. For sufficiently small 
data the field disperses. For sufficiently large data a black hole is 
formed. The question is what happens in between. This can be investigated 
by examining a one-parameter family of initial data interpolating between
the two cases. It was found by Choptuik\cite{choptuik93} that there is a
critical value of the parameter below which dispersion takes place and above
which a black hole is formed and that the mass of the black hole approaches
zero as the critical parameter value is approached. This gave rise to a
large literature where the spherical collapse of different kinds of matter
was computed numerically and various qualitative features were determined.
For a review of this see \cite{gundlach98}. In the calculations of 
\cite{rein98b} for collisionless matter it was found
that in the situations considered the black hole mass tended to a strictly
positive limit as the critical parameter was approached from above. There
are no rigorous mathematical results available on the issue of a mass gap 
for either a scalar field or collisionless matter and it is an outstanding 
challenge for mathematical relativists to change this situation.

Another aspect of Choptuik's results is the occurrence of a discretely
self-similar solution. It would seem hard to prove the existence of a
solution of this kind analytically. For other types of matter, such as
a perfect fluid with linear equation of state, the critical
solution is continuously self-similar and this looks more tractable. The
problem reduces to solving a system of singular ordinary differential 
equations subject to certain boundary conditions. A problem of this type
was solved in \cite{christodoulou94}, but the solutions produced there,
which are continuously self-similar, cannot include the Choptuik critical 
solution. In the case of a perfect fluid the existence of the critical
solution seems to be a problem which could possibly be solved in the near 
future. A good starting point for this is the work of Goliath, Nilsson
and Uggla\cite{goliath98a}, \cite{goliath98b}. These authors gave a
formulation of the problem in terms of dynamical systems and were able
to determine certain qualitative features of the solutions.

\subsection{Cylindrically symmetric solutions}\label{cylindrical} 

Solutions of the Einstein equations with cylindrical symmetry which are
asymptotically flat in all directions allowed by the symmetry represent an
interesting variation on asymptotic flatness. Since black holes are
apparently incompatible with this symmetry, one may hope to prove geodesic 
completeness of solutions under appropriate assumptions. (It would be
interesting to have a theorem making the statement about black holes
precise.) A proof of geodesic completeness has been achieved
for the Einstein vacuum equations and for the source-free Einstein-Maxwell
equations in \cite{berger95}, building on global existence theorems for
wave maps\cite{zadeh93a,zadeh93b}. For a quite different point of view 
on this question involving integrable systems see \cite{woodhouse97}.
A recent preprint of Hauser and Ernst\cite{hauser99a} also appears to
be related to this question. However, due to the great length of this text
and its reliance on many concepts unfamiliar to this author, no further
useful comments on the subject can be made here.    

\subsection{Spatially compact solutions}\label{compact}

In the context of spatially compact spacetimes it is first necessary
to ask what kind of global statements are to be expected. In a
situation where the model expands indefinitely it is natural to
pose the question whether the spacetime is causally geodesically complete
towards the future. In a situation where the model develops a singularity
either in the past or in the future one can ask what the qualitative
nature of the singularity is. It is very difficult to prove results of
this kind. As a first step one may prove a global existence theorem in
a well-chosen time coordinate. In other words, a time coordinate is chosen
which is geometrically defined and which, under ideal circumstances, will
take all values in a certain interval $(t_-,t_+)$. The aim is then to show 
that, in the maximal Cauchy development of data belonging to a certain class, 
a time coordinate of the given type exists and exhausts the expected 
interval. The first result of this kind for inhomogeneous spacetimes was
proved by Moncrief in \cite{moncrief81b}. This result concerned Gowdy 
spacetimes. These are vacuum spacetimes with two commuting Killing vectors 
acting on compact orbits. The area of the orbits defines a natural time 
coordinate. Moncrief showed that in the maximal Cauchy development 
of data given on a hypersurface of constant time, this time coordinate
takes on the maximal possible range, namely $(0,\infty)$. This result
was extended to more general vacuum spacetimes with two Killing vectors
in \cite{berger97}. Andr\'easson\cite{andreasson99} extended it in another 
direction to the case of collisionless matter in a spacetime with Gowdy 
symmetry.

Another attractive time coordinate is constant mean curvature (CMC) time.
For a general discussion of this see \cite{rendall96a}. A global existence 
theorem in this time for spacetimes with two Killing vectors and certain 
matter models (collisionless matter, wave maps) was proved in 
\cite{rendall97b}. That the choice of matter model is important for this 
result was demonstrated by a global non-existence result for dust in 
\cite{rendall97a}. As shown in \cite{isenberg98}, this leads to the
examples of spacetimes which are not covered by a CMC slicing. Related 
results have been obtained for spherical and hyperbolic symmetry 
\cite{rendall95, burnett96}. The results of \cite{rendall97b} and
\cite{andreasson99} have many analogous features and it would be
desirable to establish connections between them, since this might 
lead to results stronger than those obtained by either of the techniques
individually.

Once global existence has been proved for a preferred time coordinate, the
next step is to investigate the asymptotic behaviour of the solution as 
$t\to t_{\pm}$. There are few cases in which this has been done successfully.
Notable examples are Gowdy spacetimes \cite{chrusciel90a, isenberg90,
chrusciel90b} and solutions of the Einstein-Vlasov system with spherical
and plane symmetry\cite{rein96a}. Progress in constructing spacetimes with
prescribed singularities will be described in section \ref{prescribe}. In the 
future 
this could lead in some cases to the determination of the asymptotic behaviour
of large classes of spacetimes as the singularity is approached.

\section{Newtonian theory and special relativity}\label{newtonian}

To put the global results discussed in this article into context it is helpful
to compare with Newtonian theory and special relativity. Some of the 
theorems which have been proved in those contexts and which can offer
insight into questions in general relativity will now be reviewed. It
should be noted that even in these simpler contexts open questions abound.

\subsection{Hydrodynamics}\label{hydro} 

Solutions of the classical (compressible) Euler equations typically 
develop singularities, i.e. discontinuities of the basic fluid variables,
in finite time\cite{sideris79}. Some of the results of \cite{sideris79}
were recently generalized to the case of a relativistic 
fluid\cite{guo99a}. The proofs of the development of singularities are
by contradiction and so do not give information about what happens when
the smooth solution breaks down. One of the things which can happen is
the formation of shock waves and it is known that at least in certain
cases solutions can be extended in a physically meaningful way beyond the 
time of shock formation. The
extended solutions only satisfy the equations in the weak sense. For 
the classical Euler equations there is a well-known theorem on global
existence of classical solutions in one space dimension which goes back
to \cite{glimm65}. This has been generalized to the relativistic case.
Smoller and Temple treated the case of an isentropic fluid with linear
equation of state\cite{smoller93} while Chen analysed the cases of
polytropic equations of state\cite{chen95} and flows with variable entropy
\cite{chen97}. This means that there is now an understanding of this
question in the relativistic case similar to that available in the
classical case. 

In space dimensions higher than one there are no general global existence
theorems. For a long time there were also no uniqueness theorems for
weak solutions even in one dimension. It should be emphasized that weak
solutions can easily be shown to be non-unique unless they are required
to satisfy additional restrictions such as entropy conditions. A reasonable
aim is to find a class of weak solutions in which existence and uniqueness
hold. In the one-dimensional case this has recently been achieved by 
Bressan and collaborators (see \cite{bressan95a}, \cite{bressan95b} and 
references therein).

It would be desirable to know more about which quantities must blow up
when a singularity forms in higher dimensions. A partial answer was 
obtained for classical hydrodynamics by Chemin\cite{chemin90}. The
possibility of generalizing this to relativistic and self-gravitating
fluids was studied by Brauer\cite{brauer95}. There is one situation in which 
a smooth solution of the classical Euler equations is known to exist for all
time. This is when the initial data are small and the fluid initially 
flowing uniformly outwards. A theorem of this type has been proved by
Grassin\cite{grassin98}. There is also a global existence result due to
Guo\cite{guo98a} for an irrotational charged fluid in Newtonian physics, 
where the repulsive effect of the charge can suppress the formation of 
singularities.

A question of great practical interest for physics is that of the stability
of equilibrium stellar models. Since, as has already been pointed out, we
know so little about the global time evolution for a self-gravitating fluid
ball, even in the Newtonian case, it is not possible to say anything 
rigorous about nonlinear stability at the present time. We can, however,
make some statements about linear stability. The linear stability of a large
class of static spherically symmetric solutions of the Einstein-Euler
equations within the class of spherically symmetric perturbations has been 
proved by Makino\cite{makino98}. (Cf. also \cite{lin97} for the Newtonian
problem.) The spectral properties of the linearized
operator for general (i.e. non-spherically symmetric) perturbations in the
Newtonian problem have been studied by Beyer\cite{beyer95}.
This could perhaps provide a basis for a stability analysis, but this
has not been done. 

\subsection{Kinetic theory}

Collisionless matter is known to admit a global singularity-free
evolution in many cases. For self-gravitating collisionless matter,
which is described by the Vlasov-Poisson system, there is a general 
global existence theorem\cite{pfaffelmoser92}, \cite{lions91}. There
is also a version of this which applies to Newtonian cosmology\cite{rein94b}.
A more difficult case is that of the Vlasov-Maxwell system, which
describes charged collisionless matter. Global existence is not known
for general data in three space dimensions but has been shown
in two space dimensions\cite{glassey98a}, \cite{glassey98b} and in three
dimensions with one symmetry\cite{glassey97} or with almost spherically 
symmetric data\cite{rein90}.

The nonlinear stability of static solutions of the Vlasov-Poisson system 
describing
Newtonian self-gravitating collisionless matter has been investigated using
the energy-Casimir method. For information on this see \cite{guo99b} and its
references. 

For the classical Boltzmann equation global existence and uniqueness
of smooth solutions has been proved for homogeneous initial data and 
for data which are small or close to equilibrium. For general data with 
finite energy and entropy global existence of weak solutions (without 
uniqueness) was proved by DiPerna and Lions\cite{diperna89}. For 
information on these results and on the classical Boltzmann equation in 
general see \cite{cercignani88}, \cite{cercignani94}. Despite the 
non-uniqueness it is possible to show that all solutions tend to 
equilibrium at late times. This was first proved by Arkeryd\cite{arkeryd92} 
by non-standard analysis and then by Lions\cite{lions94} without those 
techniques. It should be noted that since the usual conservation laws for
classical solutions are not known to hold for the DiPerna-Lions solutions,
it is not possible to predict which equilibrium solution a given solution
will converge to. In the meantime analogues of several of these results for
the classical Boltzmann equation have been proved in the relativistic case. 
Global existence of weak solutions was proved in \cite{dudynski92}. Global 
existence and convergence to equilibrium for classical solutions starting 
close to equilibrium was proved in \cite{glassey93}. On the other hand global
existence of classical solutions for small initial data is not known.
Convergence to equilibrium for weak solutions with general data was proved
by Andr\'easson\cite{andreasson96}. There is still no existence and uniqueness
theorem in the literature for general spatially homogeneous solutions of the 
relativistic Boltzmann equation. (A paper claiming to prove existence and
uniqueness for solutions of the Einstein-Boltzmann system which are 
homogeneous and isotropic\cite{mucha99} contains fundamental errors.)

\section{Global existence for small data}\label{small}

An alternative to symmetry assumptions is provided by \lq small data\rq\ 
results, where solutions are studied which develop from data close to 
that for known solutions. This leads to some simplification in comparison
to the general problem, but with present techniques it is still very hard
to obtain results of this kind.

\subsection{Stability of de Sitter space}\label{desitter}

In \cite{friedrich86} Friedrich proved a result on the stability of de Sitter 
space. This concerns the Einstein vacuum equations with positive cosmological
constant. His result is as follows. Consider initial data induced by
de Sitter space on a regular Cauchy hypersurface. Then all initial
data (vacuum with positive cosmological constant) near enough to
these data in a suitable (Sobolev) topology have maximal Cauchy
developments which are geodesically complete. In fact the result gives
much more detail on the asymptotic behaviour than just this and may
be thought of as proving a form of the cosmic no hair conjecture in the
vacuum case. (This conjecture says roughly that the de Sitter solution
is an attractor for expanding cosmological models with positive
cosmological constant.) This result is proved using conformal techniques
and, in particular, the regular conformal field equations developed by
Friedrich.

There are results obtained using the regular conformal field equations
for negative or vanishing cosmological constant \cite{friedrich95,
friedrich98a} but a detailed discussion of their nature would be out of place 
here. (Cf. however section \ref{hyperboloidal}.)

\subsection{Stability of Minkowski space}\label{minkowski}

The other result on global existence for small data is that of Christodoulou
and Klainerman on the stability of Minkowski space\cite{christodoulou93b} 
The formulation of the result is close to that given in section
\ref{desitter} 
but now de Sitter space is replaced by Minkowski space. Suppose then that 
initial data are given which are asymptotically flat and sufficiently close to
those induced by Minkowski space on a hyperplane. Then Christodoulou and
Klainerman prove that the maximal Cauchy development of these data is
geodesically complete. They also provide a wealth of detail on the
asymptotic behaviour of the solutions. The proof is very long and technical.
The central tool is the Bel-Robinson tensor which plays an analogous role 
for the gravitational field to that played by the energy-momentum tensor 
for matter fields. Apart from the book of Christodoulou and Klainerman
itself some introductory material on geometric and analytic aspects of the
proof can be found in \cite{bourguignon92} and \cite{christodoulou90} 
respectively.

In the original version of the theorem initial data had to be prescribed on
all of $\R^3$. A generalization described in \cite{klainerman99} concerns
the case where data need only be prescribed on the complement of a compact 
set in $\R^3$. This means that statements can be obtained for any 
asymptotically flat spacetime where the initial matter distribution has 
compact support, provided attention is confined to a
suitable neighbourhood of infinity. The proof of the new version uses
a double null foliation instead of the foliation by spacelike hypersurfaces
previously used and leads to certain conceptual simplifications.

\subsection{Stability of the Milne model}\label{milne} 

The interior of the light cone in Minkowski space foliated by the 
spacelike hypersurfaces of constant Lorentzian distance from the origin
can be thought of as a vacuum cosmological model, sometimes known as
the Milne model. By means of a suitable discrete subgroup of the Lorentz
group it can be compactified to give a spatially compact cosmological
model. With a slight abuse of terminology the latter spacetime will also 
be referred to here as the Milne model. A proof of the stability of the
latter model by Andersson and Moncrief has been announced in 
\cite{andersson99a}. The result is that, given data for the Milne model
on a manifold obtained by compactifying a hyperboloid in Minkowski space,
the maximal Cauchy developments of nearby data are geodesically complete
in the future. Moreover the Milne model is asymptotically stable in
the sense that any other solution in this class converges towards the
Milne model in terms of suitable dimensionless variables.

The techniques used by Andersson and Moncrief are similar to those used
by Christodoulou and Klainerman. In particular, the Bel-Robinson tensor
is crucial. However their situation is much simpler than that of 
Christodoulou and Klainerman, so that the complexity of the proof is
not so great. This has to do with the fact that the fall-off of the 
fields in the Minkowksi case towards infinity is different in different 
directions, while it is uniform in the Milne case. Thus it is enough in
the latter case to always contract the Bel-Robinson tensor with the
same timelike vector when deriving energy estimates. The fact that the 
proof is simpler opens up a real possibility of generalizations, for
instance by adding different matter models.

\section{Prescribed singularities}\label{prescribe}
\subsection{Isotropic singularities}\label{isotropic}

The existence and uniqueness results discussed in this section are
motivated by Penrose's Weyl curvature hypothesis. Penrose suggests
that the initial singularity in a cosmological model should be such
that the Weyl tensor tends to zero or at least remains bounded. There
is some difficulty in capturing this by a geometric condition and it
was suggested in \cite{tod92} that a clearly formulated geometric condition
which, on an intuitive level, is closely related to the original
condition, is that the conformal structure should remain regular at the 
singularity. Singularities of this type are known as conformal or
isotropic singularities.

Consider now the Einstein equations coupled to a perfect fluid with the
radiation equation of state $p=\rho/3$. Then it has been shown \cite{newman93,
claudel98} that solutions with an isotropic singularity are determined
uniquely by certain free data given at the singularity. The data which 
can be given is, roughly speaking, half as large as in the case of a 
regular Cauchy hypersurface. The method of proof is to derive an existence
and uniqueness theorem for a suitable class of singular hyperbolic equations.
In \cite{anguige99a} this was extended to the equation of state 
$p=(\gamma-1)\rho$ for any $\gamma$ satisfying $1<\gamma\le 2$.

What happens to this theory when the fluid is replaced by a different
matter model? The study of the case of a collisionless gas of massless
particles was initiated
in \cite{anguige99b}. The equations were put into a form similar to that 
which was so useful in the fluid case and therefore likely to be
conducive to proving existence theorems. Then theorems of this kind were
proved in the homogeneous special case. These were extended to the general
(i.e. inhomogeneous) case in \cite{anguige99c}. The picture obtained for
collisionless matter is very different from that for a perfect fluid. Much
more data can be given freely at the singularity in the collisionless case.

These results mean that the problem of isotropic singularities has largely
been solved. There do, however, remain a couple of open questions. What
happens if the massless particles are replaced by massive ones? What happens
if the matter is described by the Boltzmann equation with non-trivial
collision term? Does the result in that case look more like the Vlasov
case or more like the Euler case?

\subsection{Fuchsian equations}\label{fuchsian}

The singular equations which arise in the study of isotropic singularities
are closely related to what Kichenassamy\cite{kichenassamy96a} calls
Fuchsian equations. He has developed a rather general theory of these
equations. (See \cite{kichenassamy96a}, \cite{kichenassamy96b}, 
\cite{kichenassamy96c}, and also the earlier papers \cite{baouendi77},
\cite{kichenassamy93a} and \cite{kichenassamy93b}.) In 
\cite{kichenassamy98a} this was applied to analytic
Gowdy spacetimes to construct a family of vacuum spacetimes depending on
the maximum number of free functions (for the given symmetry class) whose
singularities can be described in detail. The symmetry assumed in that paper
requires the two-surfaces orthogonal to the group orbits to be 
surface-forming (vanishing twist constants). In \cite{isenberg99a} a 
corresponding result was obtained for the class of vacuum spacetimes with 
polarized  $U(1)\times U(1)$ symmetry and non-vanishing twist.

A result of Anguige\cite{anguige99d} is of a similar type but there are 
several significant differences. He considers perfect fluid spacetimes
and can handle smooth data rather than only the analytic case. On the
other hand he assumes plane symmetry, which is stronger than Gowdy symmetry.

Related work was done earlier in a somewhat simpler context by 
Moncrief\cite{moncrief82} who showed the existence of a large class of 
analytic vacuum spacetimes with Cauchy horizons.

\section{Further results}\label{further}

\subsection{Evolution of hyperboloidal data}\label{hyperboloidal}

In section \ref{constraints} hyperboloidal initial data were mentioned. They 
can be thought 
of as generalizations of the data induced by Minkowski space on a hyperboloid.
In the case of Minkowski space the solution admits a conformal 
compactification where a conformal boundary, null infinity, can be added to
the spacetime. It can be shown that in the case of the maximal development
of hyperboloidal data a piece of null infinity can be attached to the 
spacetime. For small data, i.e. data close to that of a hyperboloid in
Minkowski space, this conformal boundary also has completeness properties
in the future allowing an additional point $i_+$ to be attached there. (See
\cite{friedrich91} and references therein for more details.) Making contact 
between hyperboloidal data and asymptotically flat initial data is much more 
difficult and there is as yet no complete picture. (An account of the
results obtained up to now is given in \cite{friedrich98a}.) If the relation
between hyperboloidal and asymptotically flat initial data could be
understood it would give a very different approach to the problem
treated by Christodoulou and Klainerman (section \ref{minkowski}).
It might well also give more detailed information on the asymptotic
behaviour of the solutions. 

\subsection{The Newtonian limit}\label{limit}

Most textbooks on general relativity discuss the fact that Newtonian
gravitational theory is the limit of general relativity as the speed
of light tends to infinity. It is a non-trivial task to give a precise
mathematical formulation of this statement. Ehlers systematized extensive 
earlier work on this problem and gave a precise definition of the Newtonian 
limit of general relativity which encodes those properties which are 
desirable on physical grounds (see \cite{ehlers91}.) Once a definition has 
been given the question remains whether this definition is compatible with
the Einstein equations in the sense that there are general families 
of solutions of the Einstein equations which have a Newtonian limit
in the sense of the chosen definition. A theorem of this kind was
proved in \cite{rendall94}, where the matter content of spacetime was assumed 
to be a collisionless gas described by the Vlasov equation. (For another
suggestion as to how this problem could be approached see 
\cite{fritelli94}.) The essential mathematical problem is that of a
family of equations depending continuously on a parameter
$\lambda$ which are hyperbolic for $\lambda\ne 0$ and degenerate for
$\lambda=0$. Because of the singular nature of the limit it is by no
means clear a priori that there are families of solutions which depend 
continuously on $\lambda$. That there is an abundant supply of families
of this kind is the result of \cite{rendall94}. Asking whether there are 
families which are $k$ times continuously differentiable in their dependence 
on $\lambda$ is related to the issue of giving a mathematical justification of
post-Newtonian approximations. The approach of \cite{rendall94} has not even 
been extended to the case $k=1$ and it would be desirable to do this. Note
however that for $k$ too large serious restrictions arise \cite{rendall92a}. 
The latter fact corresponds to the well-known divergent behaviour of higher 
order post-Newtonian approximations.  

\subsection{Newtonian cosmology}\label{cosmology}

Apart from the interest of the Newtonian limit, Newtonian gravitational
theory itself may provide interesting lessons for general relativity.
This is no less true for existence theorems than for other issues.
In this context it is also interesting to consider a slight generalization
of Newtonian theory, the Newton-Cartan theory. This allows a nice
treatment of cosmological models, which are in conflict with the (sometimes
implicit) assumption in Newtonian gravitational theory that only isolated 
systems are considered. It is also unproblematic to introduce a 
cosmological constant into the Newton-Cartan theory.

Three global existence theorems have been proved in Newtonian cosmology.
The first\cite{brauer94} is an analogue of the cosmic no hair theorem 
(cf. section \ref{desitter}) and concerns models with a positive cosmological 
constant. It asserts that homogeneous and isotropic models are nonlinearly 
stable if 
the matter is described by dust or a polytropic fluid with pressure. Thus it 
gives information about global existence and asymptotic behaviour for models
arising from small (but finite) perturbations of homogeneous and isotropic
data. The second and third results concern collisionless matter and the
case of vanishing cosmological constant. The second\cite{rein94b} says that 
data which constitute a periodic (but not necessarily small) perturbation of
a homogeneous and isotropic model which expands indefinitely give rise to
solutions which exist globally in the future. The third\cite{rein97} says 
that the homogeneous and isotropic models in Newtonian cosmology which 
correspond to a $k=-1$ Friedmann-Robertson-Walker model in general relativity 
are non-linearly stable.

\subsection{The characteristic initial value problem}\label{char}

In the standard Cauchy problem, which has been the basic set-up for all the
previous sections, initial data are given on a spacelike hypersurface.
However there is also another possibility, where data are given on one or 
more null hypersurfaces. This is the characteristic initial value problem.
It has the advantage over the Cauchy problem that the constraints reduce
to ordinary differential equations. One variant is to give initial data
on two smooth null hypersurfaces which intersect transversely in a
spacelike surface. A local existence theorem for the Einstein equations 
with an initial configuration of this type was proved in \cite{rendall90}.
Another variant is to give data on a light cone. In that case local existence
for the Einstein equations has not been proved, although it has been proved 
for a class of quasilinear hyperbolic equations which includes the reduced 
Einstein equations in harmonic coordinates\cite{Dossa97}. 

Another existence theorem which does not use the standard Cauchy problem,
and which is closely connected to the use of null hypersurfaces, concerns
the Robinson-Trautman solutions of the vacuum Einstein equations. In that
case the Einstein equations reduce to a parabolic equation. Global existence
for this equation has been proved by Chru\'sciel\cite{Chrusciel91b}.

\subsection{The initial boundary value problem}\label{ibvp}

In most applications of evolution equations in physics (and in other 
sciences) initial conditions need to be supplemented by boundary conditions.
This leads to the consideration of initial boundary value problems. It
is not so natural to consider such problems in the case of the Einstein 
equations since in that case there are no physically motivated boundary
conditions. (For instance, we do not know how to build a mirror for 
gravitational waves.) An exception is the case of fluid boundary discussed
in section \ref{freeboundary}.

For the vacuum Einstein equations it is not a priori clear that it is even
possible to find a well-posed initial boundary value problem. Thus it is
particularly interesting that Friedrich and Nagy\cite{friedrich99a} have 
been able to prove the well-posedness of certain initial boundary value
problems for the vacuum Einstein equations. Since boundary conditions 
come up quite naturally when the Einstein equations are solved numerically,
due to the need to use a finite grid, the results of \cite{friedrich99a}
are potentially important for numerical relativity. The techniques 
developed there could also play a key role in the study of the initial
value problem for fluid bodies (Cf. section \ref{freeboundary}.) 
 
\section{Acknowledgements}
I thank H\aa kan Andr\'easson and Bernd Br\"ugmann for helpful suggestions.

\bibliography{existence5}

\end{document}